\documentclass[reprint,
superscriptaddress,
 amsmath,amssymb, aps, longbibliography, nobibnotes
]{revtex4-2}
\usepackage[dvipsnames]{xcolor}
\usepackage{graphicx}
\usepackage{dcolumn}
\usepackage{bm}
\usepackage[allcolors = Blue, colorlinks = true]{hyperref}

\begin{document}

\preprint{APS/123-QED}

\title{Helical stability of double-stranded semiflexible chains with interstrand interactions}

\author{Farisan Dary}
\affiliation{Division of Physics and Applied Physics, School of Physical and Mathematical Sciences, Nanyang Technological University, Singapore 637371, Singapore}
\author{Donn Liew}
\affiliation{Division of Physics and Applied Physics, School of Physical and Mathematical Sciences, Nanyang Technological University, Singapore 637371, Singapore}
\author{Haiyi Liang}
\affiliation{Department of Modern Mechanics, University of Science and Technology of China, Hefei, Anhui Province, China}
\author{Ee Hou Yong}
\email{eehou@ntu.edu.sg}
\affiliation{Division of Physics and Applied Physics, School of Physical and Mathematical Sciences, Nanyang Technological University, Singapore 637371, Singapore}

\date{\today}
\begin{abstract}
The mechanical and structural  properties of dsDNA have been successfully described by models with varying levels of complexity and coarse-graining schemes. Prior work has characterized local stacking/twist effects and force-torque phase diagrams under external constraints. However, the role of base-pairing and torsional elasticity in global morphological transitions remain poorly characterized in the absence of external constraints. Here we investigate the delicate balance required for the strength of base-pairing interactions and the twisting energy to preserve the double-helix structure in a model made up of two semiflexible chains. We found that the model exhibits several distinct morphological phases: flat, random coil, double-helix, and the unwound double-helix. We calculate the Gauss linking number to characterize transitions between these phases.
\end{abstract}

\maketitle

\section{Introduction}
Helicity is a fundamental feature of many of the building blocks of life, including DNA, RNA, and proteins. For instance, the helical structure of double-stranded DNA (dsDNA) is essential for its biological functions such as DNA replication and genetic preservation \cite{watson1953, clegg_observing_1993}. Models with varying levels of complexity and coarse-graining schemes accurately describe the mechanical and structural properties of dsDNA. For example, the worm-like chain (WLC) model has been proven to be useful in predicting the behavior of dsDNA under external applied stresses by treating dsDNA as a semi-flexible chain \cite{ marko_bending_1994, bustamante_entropic_1994, marko_stretching_1995}. Additional modifications of the WLC model such as adding nonzero twist modulus and twist-bend coupling would explain the nonzero torsional stiffness of dsDNA \cite{skoruppa_dna_2017, skoruppa_bend_2018, gao_torsional_2021}. Alternatively, the oxDNA model treats each nucleotide in dsDNA as a rigid body with several interaction sites \cite{oulridge_dna_2010, Ouldridge2011}. Generally, a considerable number of additional energy parameters is needed to accurately model the dsDNA \cite{krajina_large-scale_2016, fosado_single_2016, gutierrez_fosado_coarse_2023}.\ \\
\indent The above-mentioned studies, among others, are mainly focused on DNA mechanics under external constraints, including local stacking/twist effects and force-torque phase diagrams. While there exist models of double-stranded polymers that exhibit distinct phases in different temperature regimes \cite{kafri2000dna, dauxois1993dynamics, badasyan2011competition}, these models do not consider independent chains that may separate upon denaturation.\ \\
\indent Here we introduce a model consisting of two semiflexible chains with interstrand interactions of strength $D$ held together under thermal fluctuations. In order to regularize the handedness, we further impose a symmetry-breaking term of strength $P$. We analyze the differences in the resulting conformation as the model parameters are varied by evaluating its Gauss linking number, and we find that our model exhibits several distinct morphological phases in the parameter space $(D,P)$. In particular, the double-helix phase is followed by a region in the phase diagram where the configuration adopts ladder-like and double-helix structure. Analyses on the Gauss linking number shows that the coexistence occurs abruptly, suggesting first-order transition.
\section{Model}
\begin{figure}[h]
    \centering
    \includegraphics{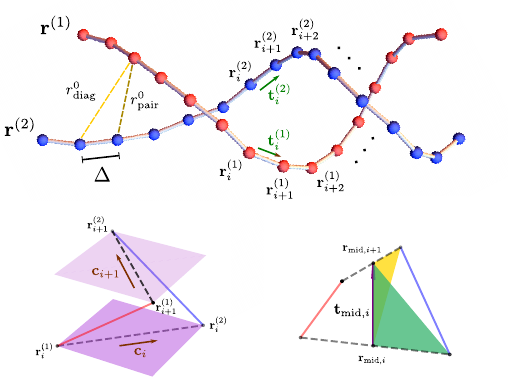}
    \caption{(a) A model of two semiflexible chains, $\mathbf{r}^{i}$ and $\mathbf{r}^{i+1}$, each consisting of vertices that are equally separated by $\Delta$. The vertices $\{\mathbf{r}_{i}\}$ are connected by tangent vector, i.e., $\mathbf{r}_{i+1} = \mathbf{r}_{i} + \Delta\cdot\mathbf{t}_{i}$. The parameters $r^{0}_{\text{diag}}$ $r^{0}_{\text{pair}}$ are the equilibrium distance of pairs that are direct opposite and diagonal opposite of each other, respectively. (b) A schematic diagram showing the counterclockwise rotation of successive corresponding vectors $\{\mathbf{c}_{i}\}$ and the triangular planes whose normals are used to regulate the twist. }
    \label{fig1}
\end{figure}
\indent Consider two discrete chains each consisting of $N$ vertices that are equally separated by a fixed distance $\Delta$. The bending flexibility of each chain is regulated by the Kratky-Porod potential \cite{kratky_rontgenuntersuchung_1949, marantan_mechanics_2018}
\begin{eqnarray}
\beta E^{(k)}_{\text{bend}} = \sum_{i=0}^{N-2}\frac{1}{g\alpha}\left(1-\cos\theta^{(k)}_{i}\right)~,
\label{eq:one}
\end{eqnarray}
where $\alpha = \Delta/\ell^{\,0}_{p}$ is a dimensionless parameter, $\theta_{i}$ denotes the bending angle between the $(i+1)$-th and $i$-th segments, and $k=1,2$ denotes each individual chain. This potential is characterized by a length scale $\ell_{p}$ and the contour length of the chain $(N-1)\Delta$. The length scale $\ell^{\,0}_{p}$ is the bare persistence length that measures the stiffness of the chain. $\ell^0_p$ is proportional to the bending rigidity $B$ with temperature dependence $\ell^{0\,}_{p}=B/k_{\text{B}}T_{0}$ \cite{bouchiat_estimating_1999, seol_elasticity_2007}. Generally, the constant $g$ has to be adjusted to match the effective Kuhn length of a continuous chain in order to reproduce the correct long-chain behavior. In this work, we set $g=1$ since the discretization parameter $\alpha$ satisfies $\alpha< 0.5$ throughout our simulations \cite{koslover2013, koslover_multiscale_2014}. \ \\ \indent The chains are held together by interstrand interactions with two specific pairings: diagonally opposite pairs and directly opposite pairs.
\begin{eqnarray}
E_{\text{diag}} = D\sum_{i = 0}^{N-2}\left[(r_{i}^{1,2}-r^{0}_{\text{diag}})^{2}+(r_{i}^{2,1}-r^{0}_{\text{diag}})^{2}\right]~
\label{eq:two}
\end{eqnarray}
where $r^{m,n}_{i} = |\mathbf{r}^{(m)}_{i}-\mathbf{r}^{(n)}_{i+1}|$ and $r^{0}_{\text{diag}}$ is the equilibrium distance. In addition, the chains are held together with specific pairings described by 
\begin{eqnarray}
 E_{\text{pair}} = \sum_{i=0}^{N-1}\frac{k}{r^{2}_{H}}\left(r_{\text{pair},i}-r^{0}_{\text{pair}}\right)^{2}~
\label{eq:three}
\end{eqnarray}
where $r_{H}$ denotes the hydrogen bond length, $r_{\text{pair},i} = |\mathbf{r}^{(1)}_{i}-\mathbf{r}^{(2)}_{i}|$ is the separation distance between the $i$-th vertex of both chains, and $r^{0}_{\text{pair}}$ is the equilibrium distance of each pair. The persistance length $\ell^{0}_{p}$ of each chain modulates the bending deformation caused by thermal fluctuations. While $E_{\text{inter}}$ and $E_\text{spring}$ provide structural stability, they are insufficient to give rise to helical configurations. This is mainly due to the fact that the corresponding vector $\mathbf{c}_{i} = \mathbf{r}^{(2)}_{i}-\mathbf{r}^{(1)}_{i}$, the vector that connects the $i$-th pair, is free to rotate in the clockwise or counterclockwise direction in order to minimize $E_{\text{inter}}$. To establish handedness, one must regulate the way both strands wind around each other by restricting the rotation of $\mathbf{c}$. In coarse-grained models where the individual strands are generated from the normal vectors along the centerline, this can be done by breaking the symmetry of the corresponding rotation vectors that connect adjacent local frames along the centerline \cite{marko_bending_1994, moroz_torsional_1997, skoruppa_bend_2018}. Since the individual chains $\mathbf{r}^{(1)}$ and $\mathbf{r}^{(2)}$ in our model are free, we do so by adding an additional energy term that breaks the handedness-symmetry such that the vector $\mathbf{c}_{i+1}$ rotates counterclockwise by a certain angle $\chi_{0}$ about the preceding vector $\mathbf{c}_{i}$:
\begin{eqnarray}
 E_{\text{twist}} = P \sum_{i=0
    }^{N-2}\left(\chi_{i}-\chi_{0}\right)^{2}~
\label{eq:four}
\end{eqnarray}
where
\[ \chi_{i}  = \begin{cases} 
      \text{acos} (\hat{\mathbf{t}}_{\text{mid},i}\cdot(\hat{\mathbf{n}}_{i}\times \hat{\mathbf{m}}_{i+1}))\\ \\
      2\pi - \text{acos} (\hat{\mathbf{t}}_{\text{mid},i}\cdot(\hat{\mathbf{n}}_{i}\times \hat{\mathbf{m}}_{i+1}))~.
   \end{cases}
\]
Here, the vector $\mathbf{n}_{i}$ describes the normal of the plane formed by $\mathbf{r}^{(2)}_{i}$, $\mathbf{r}_{\text{mid},i}$, and $\mathbf{r}_{\text{mid},i+1}$  while $\mathbf{m}_{i+1}$ describes that of $\mathbf{r}_{\text{mid},i}$, $\mathbf{r}_{\text{mid},i+1}$, and $\mathbf{r}^{(2)}_{i+1}$. The angle $\chi$ between $\mathbf{t}_{\text{mid},i} = \mathbf{r}_{\text{mid},i+1} - \mathbf{r}_{\text{mid},i}$ and the vector resulting from the product $\mathbf{n}_{i}\times \mathbf{m}_{i+1}$ determines whether $\mathbf{c}_{i}$ rotates in the clockwise or counterclockwise direction. For example, $\chi$ would be less than $\pi/2$ if the vector $\mathbf{c}$ rotates counterclockwise. Thus, we can modify the ensembles to favor right-handed (counterclockwise) turns. We set the angle $\chi_{i}$ to take the first form if $\text{sign}(\hat{\mathbf{t}}_{\text{mid},i}\cdot(\hat{\mathbf{n}}_{i}\times \hat{\mathbf{m}}_{i+1})) = 1$, and the second form if $\text{sign}(\hat{\mathbf{t}}_{\text{mid},i}\cdot(\hat{\mathbf{n}}_{i}\times \hat{\mathbf{m}}_{i+1})) = -1$. For our configuration to model relaxed dsDNA in the B-form \cite{watson1953}, we set $\chi_{0}=0.2\pi$ so that the configuration completes one turn for every $10$ pairs  .
  \begin{figure}[h]
    \centering
    \includegraphics[width=\columnwidth]{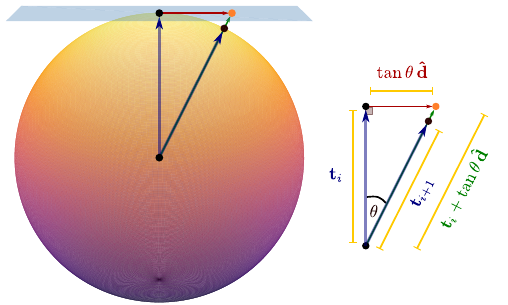}
    \caption{A schematic diagram of a chain generating algorithm. The subsequent tangent vector $\mathbf{t}_{i+1}$ is determined as the projection of the resultant vector of preceding tangent vector $\mathbf{t}_{i}$ and displacement vector $\tan\theta\,\mathbf{\hat{d}}$ onto the surface of a unit sphere. }
    \label{genchain}
\end{figure}

\paragraph*{Monte Carlo simulation.}We approximate a continuous space curve with segments of constant length $\Delta$ that connect $N$ vertices. Successive segments are generated via $\mathbf{r}_{i+1} = \mathbf{r}_{i} + \Delta\cdot\mathbf{t}_{i}$, where $\mathbf{r}_{i}$ and $\mathbf{t}_{i}$ denote the position of the $i$-th vertex and the tangent vector pointing from $\mathbf{r}_{i}$ to $\mathbf{r}_{i+1}$, respectively. Fig. \ref{genchain} shows how the successive tangent vector $\mathbf{t}_{i+1}$ can be determined by treating this problem as a random walk on the surface of a unit sphere. Given a bending angle $\theta$ between subsequent tangents $\mathbf{t}_{i+1}$ and $\mathbf{t}_{i}$, we construct the displacement vector $\mathbf{d}$ as follows. First, we choose a random unit vector $\mathbf{p}_{i}=(\sin\phi\cos\gamma,\sin\phi\sin\gamma,\cos\gamma)$, where $\phi\in[0,2\pi]$ and $\gamma\in[0,\pi]$. We then compute the displacement vector $\mathbf{d}=\mathbf{t}_{i}\times \mathbf{p}_{i}$ on the plane tangent to $\mathbf{t}_{i}$. The vector $\mathbf{t}_{i+1}$ is the projection of the displaced vector $\mathbf{t}_{i} + \tan\theta\,\mathbf{\hat{d}}$ onto the surface of the unit sphere $\mathbf{t}_{i+1} = \cos\theta\,\mathbf{t}_{i}+\sin\theta\,\mathbf{\hat{d}}$~. This method of generating a chain has been tested using the WLC model described by Eq. (\ref{eq:one}) and has been shown to yield the correct distribution of the end-to-end distance and the chain extension under applied stretching force (see Appendix \ref{wlc}). Here, we set $\mathbf{r}^{(1)}_{0}=(0,0,0)$ as a fixed reference point for the system. The total energy of the system,
\begin{eqnarray}
   E  = E^{(1)}_{\text{bend}}+ E^{(2)}_{\text{bend}} +E_{\text{pair}} + E_{\text{diag}} + E_{\text{twist}} ~
\label{eq:five}
\end{eqnarray}
is minimized via Monte Carlo with $4\times 10^{6}$ sweeps where samplings for each sweep are taken in parallel using 64 CPU-cores. We devote the first half of the simulation steps to equilibration, with configurations accepted or rejected via the Metropolis algorithm.

\paragraph*{Thermal stability of double-helix} We perform simulations at different temperatures for the double-helix configuration ($P=1$, $D=11$). We vary temperature through the dimensionless parameter $\Lambda = T/T_0$, where $T_0$ is the reference 
temperature. Since the bending rigidity $B$ is temperature-independent, varying $\Lambda$ modulates the effective persistence length $\ell_p = \ell^0_p/\Lambda$, where $\ell^0_p = B/(k_B T_0)$ is the bare persistence length. We simulate three temperatures: $\Lambda = 0.5, 0.7, 1.0$.

\section{Results}
In our simulations, each chain is made up of $N=50$ vertices that are separated by equal distance $\Delta$. The parameter values in our model are chosen in line with biologically relevant features of ssDNA and dsDNA. The arc length $\Delta$ between vertices is set to $0.64\,\text{nm}$ based on the average distance between bases in ssDNA \cite{murphy_probing_2004}. The bare persistence length $\ell^{\,0}_{p}$ of both chains is set to $2\,\text{nm}$ in order to match that of ssDNA \cite{tinland_persistence_1997, roth_measuring_2018}. Here, the parameter $k = 12\,\text{k}_{B}T/\text{nm}^{2}$, $r_{H}=0.3\,\text{nm}$, and $r^{0}_{\text{pair}}=2\,\text{nm}$ are set to replicate the hydrogen bond relative to the separation length of base pairs in dsDNA. The equilibrium distance between diagonal pairs is set to be $r^{0}_{\text{inter}} = 1.8\,\text{nm}$. The strength of the base-stacking interactions $D$ (in units $k_{\text{B}}T/\text{nm}^{2}$) and the twist $P$ (in units $k_{\text{B}}T$) are free parameters. 
  \begin{figure}[h]
    \centering
    \includegraphics[width=\columnwidth]{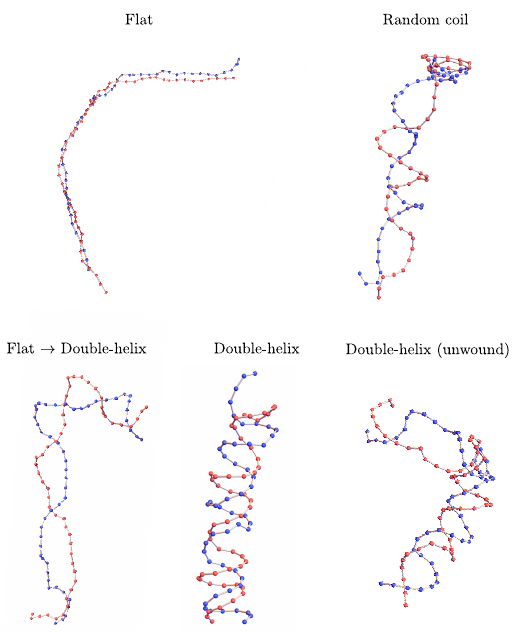}
    \caption{Morphologically distinct configurations obtained from different values of $P$ and $D$. The flat configuration corresponds to $P=D=0$, while the random coil configuration corresponds to $P=0$ and nonzero $D$. When both parameters are nonzero, the configuration adopts three different morphologies depending on the magnitude of $P$ and $D$.}
    \label{conf}
\end{figure}\\ 
\indent For $D = 0$ and $P=0$ at a fixed temperature, the chains take the form of a flat configuration. As $D$ increases with $P=0$, the chains start to wind around each other randomly since they are free to twist in any direction. Thus, the configuration consists of segments with alternating handedness. As $P$ increases, the configuration gradually adopts a double-helix configuration with an ordered handedness due to the preference of right-handed twist. The double-helix configuration start to unwind as $D$ increases further, suggesting that the morphological transitions in our model depends on the strength of the base-stacking interactions and of the energy term that breaks the handedness-symmetry. In order to quantify the difference between configurations in the parameter space $(D,P)$, we evaluate the quantities that describe the geometry of the configurations. \\
\indent The union of the corresponding vectors $\mathbf{c}$ generates a well-defined correspondence surface that is bounded by the edge curves \cite{white_calculation_1986}. In a continuous representation of our model, this correspondence surface is a ribbon whose edge curves are $\gamma_{1}=\mathbf{r}^{(1)}(s)$ and $\gamma_{2}=\mathbf{r}^{(2)}(s')$. The linking of the edge curves $\gamma_{1}$ and $ \gamma_{2}$ represents the number of times one edge curve encircles the other, which characterizes the topology of the full ribbon. The linking number is a global quantity that can be calculated from the Gauss linking integral \cite{maxwell1873}.
Here we define its discrete analog that keeps track of the link along the chains:
\begin{eqnarray}
    \text{Lk}(n) = \frac{1}{4\pi}\sum_{i=1}^{n-1}\sum_{j=1}^{n-1}\frac{\mathbf{r}_{i}^{(1)}-\mathbf{r}_{j}^{(2)}}{|\mathbf{r}_{i}^{(1)}-\mathbf{r}_{j}^{(2)}|^{3}}\cdot \biggl[\left(\mathbf{r}_{i}^{(1)}-\mathbf{r}_{i-1}^{(1)}\right)\notag \\ \times \left(\mathbf{r}_{j}^{(2)}-\mathbf{r}_{j-1}^{(2)}\right)\biggr]
\end{eqnarray}
When $n = N$, the cumulative link function $\text{Lk}(n)$ is equal to the total linking number for the entire configuration. The cumulative linking function Lk$(n)$ for morphologically distinct configurations are plotted in Fig. \ref{4}. In the transition from flat to double-helix phase and the double-helix phase, Lk$(n)$ increases linearly. However, the fully formed double-helix achieves higher total linking values. In the unwound double-helix phase, the total linking is reduced compared to the double helix phase because a portion of the double-helix unwinds and forms a ladder-like configuration, evident from nonlinear trends in Lk$(n)$. The linking remains small in the flat phase since bending and twisting are rare. In the random coil phase, Lk$(n)$ may become negative since there is no preferred twisting direction due to the unbroken handedness symmetry. 
\begin{figure}[h]
    \centering
    \includegraphics[width=\columnwidth]{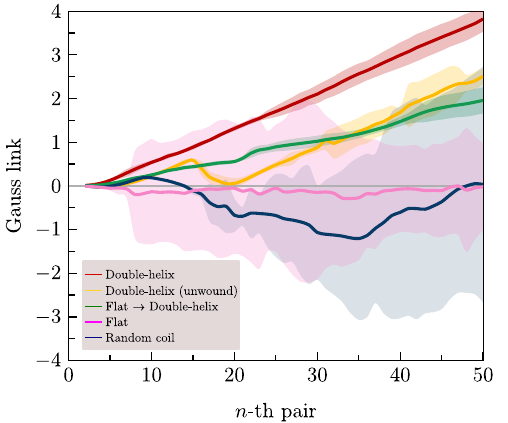}
    \caption{The cumulative link function Lk$(n)$ of configurations in different morphological phases. The link in double-helix phase (red) and during the transition (green) from flat to double-helix is linearly increasing along the configuration. In unwound double-helix phase, the nonlinear trends in Lk$(n)$ corresponds to a ladder-like configuration. The link is relatively small in the flat phase (lavender) since the  configuration rarely bends and twists. In the random coil phase (blue), Lk$(n)$ fluctuates between positive and negative link since the configuration is allowed to twist freely in any direction since the handedness-symmetry is unbroken. }
    \label{4}
\end{figure}\\
\indent In the case of $P=0$ and $D=10$, the average of Lk$(n)$ is zero since the handedness-symmetry has not been broken. As shown in Fig. \ref{5}(a), Lk$(n)\neq0$ since the configuration abruptly adopts an unwound double-helix as $P$ increases with fixed $D$. The sharp increase in the link represents the disappearance of ladder-like configuration, resulting in the instantaneous transformation to a full double-helix. In the case of $P=1$ and $D=0$, we find that small amount of twisting and bending has already contributed to Lk$(n)$. Fig. \ref{5}(b) shows how Lk$(n)$ gradually increases as the configuration steadily transforms into a double-helix as $D$ increases with fixed $P$. The sharp decrease in Lk$(n)$ that follows as $D$ increases further signifies the abrupt unwinding of the double-helix configuration. In both cases, the total link Lk$(N)$ is not a suitable order parameter to capture the abrupt transition near the critical point. The unsuitability is due to the bending and twisting of the unwound portion of the configuration, thus contributing to variations in Lk$(N)$. Since our configuration is relatively short, the total link is sensitive to these variations thus rendering it unreliable near the critical points. Instead, we use the ratio $N_u/N$ as the order parameter, where $N_u$ is the number of unwound pairs and N is the total number of pairs. $N_u/N$ is robust because it is unaffected by bending and twisting variations in the unwound regions. Since $N_u = N - N_h$, where $N_h$ is the number of pairs in double-helix configuration, we determine $N_h$ by analyzing the cumulative linking profile Lk$(n)$. We define a segment $i$ as helical if its local linking density $\Delta$Lk$=$Lk$(i+1)-$Lk$(i)$ exceeds a threshold of $\delta=0.05$. Segments below this threshold are classified as unwound.
\indent Consider the case $P=1$ where the critical point $D_{c}$ lies within $14\leq D_{c}\leq 15$. As shown in Fig. \ref{5}(c), the proliferation of unwound portion is captured clearly as $D$ passes through $D_c$. This is indeed reminiscent of a first-order phase transition. The unwound portion evidently grows in size as $D$ increases, and $N_u/N\rightarrow1$ as $D\rightarrow\infty$. 
As $N_u/N\rightarrow1$, the system approaches a fully unwound state consisting primarily of ladder-like segments. The overall morphology is therefore closely related to the energy distribution within the configuration.
\paragraph*{Phase diagram.} Morphologically distinct configurations can be characterized based on several criterion. We collect and identify the morphology of configurations taken from different values of $P$ and $D$ in the phase diagram as plotted in Fig. \ref{5}(d). The transition from flat to double-helix phase is smooth, with the transition region decreasing as $P$ gets larger. At higher P values, the system has a stronger tendency to form right-handed twists, leading to the formation of double-helix at lower base-stacking interaction strength $D$. The phase transition from the random coil phase to phases with ordered handedness is abrupt due to the handedness-symmetry breaking. 
\begin{figure*}[htb!]
    \centering
    \includegraphics[width=\textwidth]{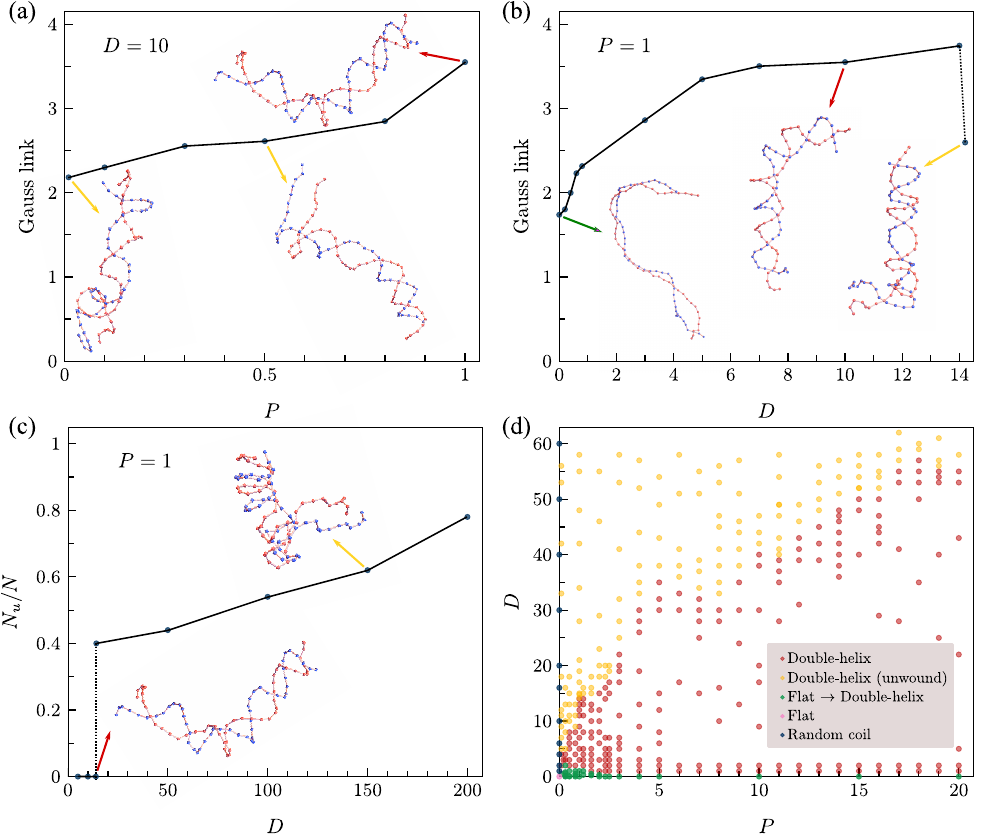}
    \caption{Morphological phases of the model with different values of $P$ and $D$: (a) When $D=10$ is fixed, the configuration becomes right-handed spontaneously when $P$ is no longer zero. The configuration is a partially unwound double-helix (yellow $\to$), which eventually becomes full double-helix (red $\to$) as $P$ increases further. (b) When $P=1$ fixed, the configuration with $D=0$ has nonzero linking number (green $\to$), which steadily (signified by the gradual rise in the Gauss link) becomes a double-helix as as $D$ increases. As $D$ increases further, the double-helix abruptly unwound. (c) Similar to (b) but instead of Gauss link, the quantity $N_{u}/N$ which represents the fraction of unwound segments is used to describe the transition as $D$ increases. The sharp increase in $N_{u}/N$ as $D$ passes the critical point $D_{c}$ is a reminiscent of a first-order phase transition. (d) Phase diagram constructed from the previous criteria, showing regions corresponding to different morphological phases. The flat phase is a single point $(0,0)$ in the parameter space, and the region corresponding to the transition from flat to double-helix becomes smaller as $P$ gets larger. The double-helix structure becomes more stable as $P$ gets larger, as shown from the increase in its phase boundary with the region corresponding to unwound double helix phase.}
    \label{5}
\end{figure*}
\section{Discussion}
We calculate Tw, Wr, and Lk to analyze the geometrical differences between the double-helix and unwound double-helix phases. Twist Tw $=\frac{1}{2\pi}\int dU(t\times U)$ measures the cross section rotation rate \cite{white_calculation_1986}, where $U$ is a vector normal to tangent $t$. Our surface uses correspondence vectors $\hat{c}$ between chains. Since $\hat{c}$ rotates freely, a consistent $U$ cannot be defined. Thus, the calculation of Tw is ambiguous since it would require us to find the vector $\mathbf{U}$ in the plane spanned by $\mathbf{\hat{c}}$ and the tangent vector $\mathbf{t}$ at every step in the simulation. Since the pairing interactions maintain small interstrand separation, the ribbon writhe may be approximated as the centerline writhe $\mathbf{r}^{\text{mid}} =(\mathbf{r}^{(1)}+\mathbf{r}^{(2)})/2$. We define a cumulative writhe function $W(n)$ as:
\begin{eqnarray}
  W(n) = -\frac{1}{2\pi}\sum_{i=2}^{n-2}\sum_{j<i}\Omega_{ij}~,
\end{eqnarray}
where $\Omega_{ij}$ is the Gauss integral computed between centerline segments $i$ and $j$, given by vectors $\mathbf{r}^{\text{mid}}_{i+1}-\mathbf{r}^{\text{mid}}_{i}$ and $\mathbf{r}^{\text{mid}}_{j+1}-\mathbf{r}^{\text{mid}}_{j}$ respectively, for $j<i$ \cite{klenin_computation_2000}. Since twisting energy $E_{\text{twist}}$ is orientable, we calculate Tw using the Călugăreanu–White–Fuller theorem, Tw$(n)=\text{Lk}(n)-W(n)$ \cite{moroz_entropic_1998, vologodskii1994conformational}. From Tw$(n)=\text{Lk}(n)-W(n)$, we decompose linking into two contributions: $W(n)$ from the centerline's 3D geometry and Tw$(n)$ from local twisting. At $n = N$, we obtain the total Tw and Wr for the entire configuration. We note that configurations with knotted centerlines are rare in our simulations (see Appendix \ref{knot}). Additionally, E$_\text{pair}$ and E$_\text{diag}$ prevent configurations where an unknotted centerline has boundary curves forming (2,2k)-torus links \cite{orlandini_statistical_2007}.



\begin{figure}[h]
    \centering
    \includegraphics[width=\columnwidth]{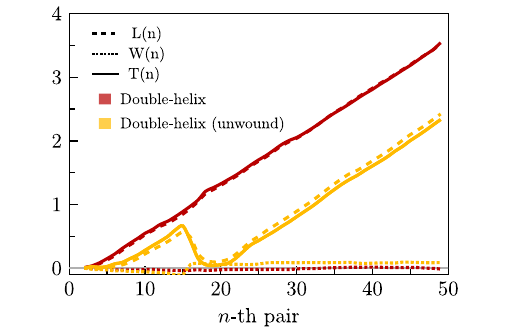}
    \caption{The cumulative functions  Lk$(n)$, Tw$(n)$, and W$(n)$ of a configuration in double-helix phase and unwound double-helix phase. The nonlinear trends of Lk$(n)$ and Tw$(n)$ in unwound double-helix phase  indicate portions where the double-helix structure is lost. The small increase in $W(n)$ shortly before the linear  increase in Lk$(n)$ and Tw$(n)$ indicates that the ladder-like and double-helix structure are bridged by bending. The similarity between Tw$(n)$ in both phases suggests that the double-helix structure of both configurations, share the same rate of twist despite  being in morphologically distinct phases.}
    \label{lkdiff}
\end{figure}
\indent Fig. \ref{lkdiff} shows  Lk$(n)$, Tw$(n)$, and $W(n)$ for parameter space $(P=1, D=14)$ (double-helix phase) and $(P=1, D=15)$ (unwound double-helix phase). $W(n)$ remains small in both cases because the configurations are relatively short. However, $W(n)$ increases before Tw$(n)$ and Lk$(n)$ begin increasing linearly in the unwound phase. This increase reflects greater centerline nonplanarity due to bending, as shown in Fig. \ref{conf}. Tw$(n)$ increases monotonically in both phases, consistent with right-handed twisting (positive twist). Importantly, the rate of change of Tw$(n)$ are similar in both phases, indicating that the local twisting rate remains constant in helical regions. This suggests that when $D$ is large, forming a semi-stable state with coexisting ladder-like and helical segments costs less energy than maintaining a fully helical structure with higher twist. We verify this by calculating $\epsilon$, the average energy contribution per segment:
\begin{eqnarray}
    \varepsilon = \frac{E^{(1)}_{\text{bend}}+E^{(2)}_{\text{bend}}}{N-2} + \frac{E_{\text{pair}}}{N} + \frac{E_{\text{diag}}}{2(N-1)} + \frac{E_{\text{twist}}}{N-1}~
\end{eqnarray}
where the energy terms $E$ are defined in eq (\ref{eq:one})--(\ref{eq:five}), with each energy term normalized by its number of contributions: $N-2$ bending angles per chain, $N$ paired vertices, $2(N-1)$ diagonal pairs, and $N-1$ twist angles respectively.

For instance, the configuration with $(P=1,D=11)$ adopts a full double-helix configuration with $\varepsilon=1.59\,k_{B}T$, while for $(P=1,D=15)$ the configuration is in a semi-stable state with $\varepsilon = 1.66\,k_{B}T$. To show that unwinding is energetically favorable, we calculate the energy of the D=11 helical geometry using D=15 parameters, obtaining $\varepsilon$ = 2.48 $k_{B}T$. Since the actual mixed state at D=15 has lower energy $(1.66 k_{B}T < 2.48 k_{B}T)$, the system reduces its energy by forming ladder-like segments that lower bending costs despite stronger diagonal interactions.
 
\begin{figure}[h]
    \centering
    \includegraphics{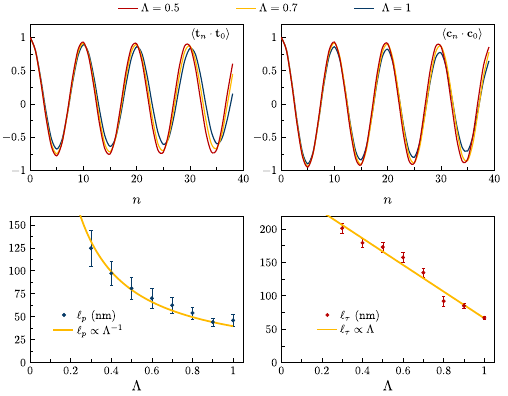}
    \caption{The correlation functions $\langle\mathbf{t}_{n}\cdot\mathbf{t}_{0}\rangle$  (top left) and $\langle\mathbf{c}_{n}\cdot\mathbf{c}_{0}\rangle$  (top right) exhibit exponential oscillatory decay at different temperature $\Lambda$.  Temperature dependence of bending persistence length $\ell_p$  (bottom left) and torsional persistence length $\ell_{\tau}$  (bottom right) are obtained by fitting $\langle\mathbf{t}_{n}\cdot\mathbf{t}_{0}\rangle = e^{-s/\ell_{p}}\cos(\lambda_{p} s)$ and $\langle\mathbf{c}_{n}\cdot\mathbf{c}_{0}\rangle = e^{-s/\ell_{\tau}}\cos(\lambda_\tau s)$, respectively. $\ell_{p}$ increases as $\Lambda^{-1}$ when $\Lambda$ is lowered, while $\ell_{\tau}$ increases linearly with decreasing $\Lambda$.}
    \label{fig2}
\end{figure}

 Since thermal fluctuations  are known to instigate conformational changes in dsDNA \cite{depew_conformational_1975, duguet_helical_1993, kriegel_temperature_2018}, we investigate the stability of our double-helix configuration at  varying temperatures via $\Lambda=T/T_0$. In order to quantify structural regularity, we fit the correlation functions to exponentially damped oscillations $\langle\mathbf{t}_n\cdot\mathbf{t}_0\rangle = e^{-s/\ell_p}\cos(\lambda_p s)$ and $\langle\mathbf{c}_n\cdot\mathbf{c}_0\rangle = e^{-s/\ell_\tau}\cos(\lambda_\tau s)$, where $\ell_\tau$ is the torsional persistence length measuring the regularity of the helical pitch. In Fig. \ref{fig2}, we plot the correlation functions $\langle\mathbf{t}_{0}\cdot\mathbf{t}_{n}\rangle,$ $\langle\mathbf{c}_{n}\cdot\mathbf{c}_{0}\rangle$ at $\Lambda=0.5$, $\Lambda=0.7$, and $\Lambda=1$  and persistence lengths $\ell_p, \ell_\tau$ for $\Lambda=[0.3,~1]$. Our simulation results show that the the tangent-tangent correlation $\langle\mathbf{t}_{0}\cdot\mathbf{t}_{n}\rangle$ of both strands exhibit oscillatory behavior unlike  the WLC which decays purely exponentially at any temperature \cite{giomi2010, liverpool1998statistical}. The oscillatory behavior of $\langle\mathbf{t}_{n}\cdot\mathbf{t}_{0}\rangle$ has been observed in single  helical ribbons \cite{giomi2010,yong2022}. The  corresponding vector correlation $\langle\mathbf{c}_{n}\cdot\mathbf{c}_{0}\rangle$ also displays oscillatory behavior, suggesting the  structural regularity of the double-helix model.
\\
\indent We  fit these correlation functions with  exponentially damped oscillations:
\begin{eqnarray}
    \langle\mathbf{t}_{n}\cdot\mathbf{t}_{0}\rangle = e^{-s/\ell_{p}}\cos(\lambda_{p} s)~, \\ \langle\mathbf{c}_{n}\cdot\mathbf{c}_{0}\rangle = e^{-s/\ell_{\tau}}\cos(\lambda_\tau s)
\end{eqnarray}
where $s = n\Delta$. Here the wavenumbers $\lambda_{p}$ and $\lambda_{\tau}$ are  related to the helical pitch. Bending persistence length $\ell_{p}$ increases as $\Lambda^{-1}$ when $\Lambda$ is lowered, while the torsional persistence length $\ell_{\tau}$ increases linearly with decreasing $\Lambda$. The effective persistence length $\ell_{p}$ at $\Lambda = 1$ is found to be approximately $20$ times larger than $\ell^{0}_{p}$.

\indent The effective persistence length $\ell_p$ of the double-helix at $\Lambda = 1$, measured from tangent-tangent correlations, is approximately 20 times larger than the bare persistence length $\ell_p^0 = 2$ nm of individual chains. This enhancement reflects geometric stiffening from the coupled helical structure, consistent with the ratio between double-stranded DNA $(\ell_p \sim 50 \text{ nm})$ and single-stranded DNA $(\ell_p \sim 2-3 \text{ nm})$ \cite{roth_measuring_2018, tinland_persistence_1997}. The stiffening arises primarily from base-stacking interactions in the helical geometry \cite{mills2004origin}, rather than from simply adding the stiffnesses of two independent chains \cite{liverpool1998statistical}.
\section{Conclusion}
In summary, we found that our model exhibits distinct morphological phases which can be characterized from its Gauss linking number. Moreover, the stability of a double-helix structure in our model has been found to be dependent on the strength of its individual energy terms. In particular, the double-helix could become unstable which causes it to partially unwound spontaneously when the diagonal base-stacking interactions is much stronger than its twisting rigidity, a reminiscent of a  first-order phase transition. This can be understood as a mechanism for the configuration to minimize its energy by reducing the bending of its segments. The fraction of the unwound segments has been shown to be increasing with the strength of  interstrand stacking interactions, and the double-helix eventually unwounds entirely as its diagonal  stacking interactions become much stronger compared to its twisting rigidity.

\section{Acknowledgement}

F.D, D.L., and E.H.Y. acknowledge support from Singapore Ministry of Education through the Academic Research Fund Tier 1 (RG140/22) and Academic Research Fund Tier 2 (MOE-T2EP50223-0014). The computational work for this article was partially performed on resources of the National Supercomputing Centre, Singapore (https://www.nscc.sg). H.L. acknowledges support from the National Key Research and Development Program of China (Grant No. 2025YFA0922902).

\appendix
\section{Statistics of worm-like chain}
\label{wlc}
The theoretical values of the end-to-end distance as a function of the total number of monomers $N$ is given by 
\begin{eqnarray}
    \frac{\sqrt{\langle R^{2}\rangle}}{\Delta} = \sqrt{\frac{2N}{\alpha}\biggl(1-\frac{1}{N\alpha}\left(1-e^{-N\alpha}\right)\biggr)}~
\end{eqnarray}
While the behavior of its extension $z/L$ under stretching force $f = F\ell_{p}/(k_{\text{B}}T)$ can be obtained via path integral approach as outlined in Ref. \cite{bouchiat_estimating_1999}. 

\section{Knotting statistics}
\label{knot}
 \indent The topology of dsDNA is usually understood based on the knotting of the central axis of a closed and circular dsDNA molecules. Since our chain configuration is open, we demonstrate the rarity of knottings in our simulations by identifying the knot topology of the middle curve $\mathbf{r}^{\text{mid}}=\left(\mathbf{r}^{(1)}+\mathbf{r}^{(2)}\right)/2$ with minimally interfering chain closure algorithm \cite{tubiana2011, tubiana2018}. We analyze the knot topology at every 500 steps after equilibration and found that the knotting probability is around $0.5\%$.

\section*{Data availability}
\noindent All code and relevant data can be found on \href{https://github.com/donn-liew/dwlc}{Github} at \href{https://github.com/donn-liew/dwlc}{https://github.com/donn-liew/dwlc}.

\newpage
\bibliography{apssamp}

\end{document}